\documentstyle [12pt] {article}
\input epsf

\newcommand{\beq}{\begin{equation}}
\newcommand{\eeq}{\end{equation}}
\newcommand{\beqs}{\begin{eqnarray}}
\newcommand{\eeqs}{\end{eqnarray}}

\begin{document}

\def\thefootnote{\fnsymbol{footnote}}
\baselineskip 6.0mm

\begin{flushright}
\begin{tabular}{l}
ITP-SB-96-19    \\
May, 1996
\end{tabular}
\end{flushright}

\vspace{4mm}
\begin{center}
{\bf Complex-Temperature Phase Diagram of the 1D $Z_6$ Clock Model}

\vspace{4mm}

{\bf and its Connection with Higher-Dimensional Models}

\vspace{12mm}

\setcounter{footnote}{0}
Victor Matveev\footnote{email: vmatveev@insti.physics.sunysb.edu}
\setcounter{footnote}{6}
and Robert Shrock\footnote{email: shrock@insti.physics.sunysb.edu}

\vspace{6mm}
Institute for Theoretical Physics  \\
State University of New York       \\
Stony Brook, N. Y. 11794-3840  \\

\vspace{12mm}

{\bf Abstract}
\end{center}
   We determine the exact complex-temperature (CT) phase diagram of the
1D $Z_6$ clock model. This is of interest because it is the first exactly
solved system with a CT phase boundary exhibiting a finite-$K$ intersection 
point where an odd number of curves (namely, three) meet, and yields a 
deeper insight into this phenomenon. Such intersection points occur in the 
3D spin 1/2 Ising model and appear to occur in the 2D spin 1 Ising model.  
Further, extending our earlier work on the higher-spin Ising model, we point 
out an intriguing connection between the CT phase diagrams for the 1D and 2D 
$Z_6$ clock models. 

\vspace{16mm}

\pagestyle{empty}
\newpage

\pagestyle{plain}
\pagenumbering{arabic}
\renewcommand{\thefootnote}{\arabic{footnote}}
\setcounter{footnote}{0}

   We report here a determination of the exact complex-temperature phase 
diagram for the 1D $Z_6$ clock model and show how the results give a deeper 
insight into certain features of higher-dimensional models.  The idea of 
generalizing a variable, on which the free energy depends, from real physical 
values to complex values was pioneered by Yang and Lee \cite{yl}, who carried
this out for the external magnetic field and proved a celebrated circle 
theorem on the complex-field zeros of the Ising model partition function.  
The generalization of temperature to complex values, performed first 
for the Ising model \cite{fisher}, has also been quite fruitful, since 
it enables one to understand better the behavior of various thermodynamic 
quantities as analytic functions of complex temperature (CT) and to see how 
physical phases of a given model generalize to regions in CT variables.  
Indeed, a knowledge of the complex-temperature phase diagram of a model for 
which the free energy is not known provides further constraints to guide 
progress toward an exact solution.  A classic example relevant here is the 
3D Ising model; the main purpose of the present work is to present some
exact results for a 1D model which elucidate an important feature of the
complex-temperature phase diagram of the 3D Ising model. Some early works on CT
singularities include Refs. \cite{katsura}-\cite{ks} studying partition
function zeros, and Ref. \cite{dg}, motivated by the effect of these 
singularities on low-temperature series.  The continuous locus
of points in the complex-temperature plane where the free energy is
non-analytic is denoted ${\cal B}$. For the spin models of interest here, with
isotropic, nearest-neighbor spin-spin couplings, this is a 1-dimensional 
curve (including possible line segments).  Parts of ${\cal B}$ form 
boundaries of various regions, some of which are complex-temperature 
extensions of physical phases and some of which may have no overlap with any 
physical phase.  ${\cal B}$ may also include arcs or line segments which 
protrude into, and terminate in, certain CT phases.

    In general, phase boundaries of physical systems include intersection
points where different parts of the boundaries meet. Examples 
include (i) the triple point in the $(T,p)$ phase diagram for a substance 
like argon, water, etc. where gas, liquid, and solid phases all coexist, and 
(ii) the two triple points forming
the ends of the $\lambda$ line in the phase diagram for helium, where, at the
lower end of this line, the gas, normal fluid and superfluid coexist, and, at
the upper end, the solid coexists with the normal fluid and superfluid. 
Similarly, the complex-temperature phase boundary ${\cal B}$ of a spin system
may include intersection points at which different curves contained in ${\cal
B}$ meet.  We denote the number of
curves meeting at such an intersection point as $n_c$.  Often these can be
grouped into pairs, such that two curves meet with equal tangents at the
intersection point; these are then regarded as a single branch of a curve
passing through this point.  In the terminology of algebraic geometry
\cite{alg}, a singular point of an algebraic curve is a multiple (intersection)
point of index $m$ if $m$ branches of the curve pass through this point.  This
is thus an intersection point with $n_c=2m$ curves meeting, in $m$ pairs with
$n_t=m$ equal tangents.  For the 2D Ising model on the (homopolygonal) square, 
triangular, and honeycomb lattices, and also on the heteropolygonal 
$3 \cdot 6 \cdot 3 \cdot 6$ (kagom\'e) and $3 \cdot 12 \cdot 12$ lattices, the
CT phase boundaries ${\cal B}$ are algebraic curves and their intersection 
points always have $n_c=4$ and $n_t=2$ so that they are regular multiple points
of index $m=2$ (see Tables 2,3 in Ref. \cite{cmo}).  This is also true of the
Ising model in a nonzero external field satisfying the Lee-Yang condition 
$\beta H = i \pi/2$; the CT phase boundaries ${\cal B}$ on the square,
triangular, honeycomb, and $3 \cdot 12 \cdot 12$ lattices all exhibit
intersection points with $n_c=4$, $n_t=2$ which are thus again regular multiple
points of index $m=2$ \cite{ih}.  In all of these cases, the two branches of
the curves cross with an angle $\theta_c=\pi/2$, i.e., orthogonally. 

   However, in other models, it appears that the respective phase boundaries
${\cal B}$ have intersections at finite complex-temperature points with 
the odd  $n_c=3$.  In contrast to the 2D Ising model results noted 
above, which are extracted (in Refs. \cite{fisher,abe,cmo,chisq,chith}) from 
exact solutions (\cite{ons} for square lattice; reviewed for general lattices 
in Ref. \cite{revs}), up to the present time, to our knowledge, none of the 
models which appear to have finite-$K$ intersection points with odd $n_c$ has 
been exactly solved (here, $K=J/(k_BT)$). 
Instead, the occurrence of such points has been inferred
from inspection of CT zeros of the partition function 
calculated on finite lattices. For example, for
the (spin 1/2) Ising model on the simple cubic lattice, such zeros 
\cite{oksk,p} suggest an intersection point at $u_{1/2} \simeq -0.5$, where 
$u_s=e^{-K/s^2}$.  At this point, a component of ${\cal B}$ crossing the 
negative real $u_{1/2}$ axis vertically meets a line segment lying on this
axis in a T intersection, so that $n_c=3$, $n_t=2$.  Since the lattice is 
bipartite, there is a similar intersection point at the inverse position, 
$u_{1/2} \simeq -2$. 
Similarly, in our studies of the spin 1 Ising model on the square lattice we 
have found, for the largest lattice sizes, indications of $n_c=3$ 
intersection points at $u_1 \simeq 0.1 + 0.8i$, its inverse, and their 
complex-conjugates \cite{hs}.  
To gain further insight into this feature of complex-temperature
phase boundaries, one is motivated to search for an exactly solvable model 
which exhibits this behavior.  We have succeeded in this; the 1D $Z_6$ clock
model provides to our knowledge the first exactly solvable model with a
finite-$K$ intersection point having odd $n_c$, and we analyze it here. 

   The $Z_N$ clock model is defined at temperature $T$ 
by the partition function $Z=\sum_{\{ \ell_n \} }e^{-\beta{\cal H}}$ with 
$\beta = (k_BT)^{-1}$ and 
\beq
{\cal H} = \sum_{\langle nn' \rangle } E(-\ell_n+\ell_{n'})
\label{ham}
\eeq
where the interaction energy is 
\beq
E(-\ell_n + \ell_{n'}) = -J\cos (2 \pi(-\ell_n + \ell_{n'})/N) 
\label{e}
\eeq
the site variable is $\ell_n=0,1,...N-1$, and $\langle nn' \rangle$ denotes
nearest-neighbor pairs.  Equivalently, $E=-J{\bf S}_n \cdot {\bf S}_{n'}$, 
where the angles of the (classical) spins take on the discrete values 
${\bf S}_n=(\cos \theta_n, \sin \theta_n)$ with $\theta_n=2\pi \ell_n/N$. 
We use the notation $K = \beta J$ and $u=e^{-K/2}$.  We consider the case 
$N=6$ here because it provides a simple exactly solvable example of an 
intersection point with odd $n_c$ on a complex-temperature phase boundary 
${\cal B}$. The (reduced) free energy is 
$f = -\beta F = \lim_{N_s \to \infty} N_s^{-1} \ln Z$ in the thermodynamic
limit.  

   For $d=1$ dimension, one can solve this model exactly, e.g., by transfer 
matrix methods. One has 
\beq
Z = Tr({\cal T}^{N_s}) = \sum_{j=1}^6 \lambda_j^{N_s}
\label{zeq}
\eeq
where the $\lambda_j$, $j=1,...6$ denote the eigenvalues of the 
transfer matrix ${\cal T}$ defined by 
${\cal T}_{nn'}= \langle \ell_n|\exp(-\beta E(\ell_n, \ell_{n'}))|\ell_{n'} 
\rangle$, and we use periodic boundary conditions.  It is 
convenient to analyze the phase diagram in the $u$ plane. For physical
temperature, phase transitions are associated with degeneracy of leading 
eigenvalues \cite{al}. 
There is an obvious generalization of this to the case of complex
temperature: in a given region of $u$, the eigenvalue of ${\cal T}$ 
which has maximal magnitude, $\lambda_{max}$,  gives the 
dominant contribution to $Z$ and hence, in the thermodynamic 
limit, $f$ receives a contribution only from $\lambda_{max}$: 
$f=\ln ( \lambda_{max})$.  For complex $K$, $f$ is, in general, 
also complex.  The CT phase boundaries are determined by the degeneracy, in
magnitude, of leading eigenvalues of ${\cal T}$.  
As one moves from a region with one dominant (i.e., leading) eigenvalue 
$\lambda_{max}$ to a region in which a different eigenvalue $\lambda_{max}'$ 
dominates, there is a non-analyticity in $f$ as it switches from 
$f=\ln (\lambda_{max})$ to $f=\ln (\lambda_{max}')$.  The boundaries of these
regions are defined by the degeneracy condition among dominant eigenvalues, 
$|\lambda_{max}|=|\lambda_{max}'|$.  These form curves in the $u$ plane.  
Note that the free energy and the conditions for CT phase boundaries are the 
same if some eigenvalues occur with finite multiplicity greater than one.
To see this, assume $\lambda_j$ occurs $n_j$ times, where $n_j$ is finite. If 
$\lambda_j$ is nonleading, the result is obvious; if $\lambda_j$ is leading,
the result follows because $\lim_{N_s \to \infty}N_s^{-1} 
\ln (n_j\lambda_j^{N_s}) = \ln \lambda_j$ for any finite (nonzero) $n_j$.  
This is relevant here because two of the eigenvalues of ${\cal T}$ occur with 
multiplicity 2. 

   Of course, a 1D spin model with finite-range interactions has no 
non-analyticities for any (finite) value of $K$, so that, in particular,
the 1D $Z_6$ model is analytic along the positive real $u$ axis. 
Because of the invariance of $Z$ on a bipartite lattice under the
transformation ${\bf S}_e \to {\bf S}_e$, ${\bf S}_o \to -{\bf S}_o$, $J \to
-J$, where $e$ and $o$ denote sites on even and odd sublattices, it follows 
that 
\beq
u \to 1/u \ \ \Longrightarrow {\cal B} \ \ {\rm invariant} 
\label{usym}
\eeq
(For a finite 1D lattice with periodic boundary conditions, we preserve this 
invariance by using even $N_s$.) Further, 
since the $\lambda_j$ are analytic functions of $u$, whence 
$\lambda_j(u_s^*)=\lambda_j(u_s)^*$, it follows that the solutions to 
the degeneracy equations defining the boundaries between different phases,
$|\lambda_i|=|\lambda_j|$, are invariant under $u \to u^*$.  
Hence, ${\cal B}$ is invariant under $u \to u^*$.

Although the model has a physical phase structure consisting only of the 
$Z_6$-symmetric, disordered phase, its complex-temperature phase diagram
is nontrivial and exhibits a number of interesting features. An important 
property of this model is that the interaction energies $E(\Delta \ell)$, where
$\Delta \ell = \ell_n - \ell_{n'}$, are all rational multiples of each other: 
$E(\Delta)/J=-1$, $-1/2$, 1/2, and 1 for $\Delta \ell = 0$, $\pm
1$, $\pm 2$, and $\pm 3$, respectively, with 
$E(6-\Delta \ell)=E(\Delta \ell)$. Hence, all of the Boltzmann weights are
powers of $u=e^{-K/2}$.  The transfer matrix ${\cal T} = 
\langle \ell_n|e^{-\beta {\cal H}}|\ell_{n'}\rangle$ has the Toeplitz form
\beq
{\cal T} =  u^{-2}\left (\begin{array}{cccccc}
           1   & u   & u^3 & u^4 & u^3 & u   \\
           u   & 1   & u   & u^3 & u^4 & u^3 \\
           u^3 & u   & 1   & u   & u^3 & u^4 \\
           u^4 & u^3 & u   & 1   & u   & u^3 \\
           u^3 & u^4 & u^3 & u   & 1   & u   \\
           u   & u^3 & u^4 & u^3 & u   & 1   \end{array}   \right  )
\label{tmatrix}
\eeq
The eigenvalues are 
\beq
\lambda_1 = u^{-2} - 2u^{-1} + 2u - u^2 = u^{-2}(1+u)(1-u)^3
\label{lam1}
\eeq
\beq
\lambda_2 = u^{-2} + 2u^{-1} + 2u + u^2 = 
u^{-2}\Bigl ( 1+(1+\sqrt{3})u+u^2 \Bigr ) \Bigl (1+(1-\sqrt{3})u+u^2 \Bigr )
\label{lam2}
\eeq
\beq
\lambda_3 = \lambda_4 = u^{-2} + u^{-1} - u - u^2 = u^{-2}(1-u)(1+u)(1+u+u^2)
\label{lam34}
\eeq
\beq
\lambda_5 = \lambda_6 = u^{-2} - u^{-1} - u + u^2 = 
             u^{-2}(1-u)^2(1+u+u^2)
\label{lam56}
\eeq
and we set $\lambda_{34} \equiv \lambda_3 = \lambda_4$ and 
$\lambda_{56} \equiv \lambda_5 = \lambda_6$. 

\begin{figure}
\epsfxsize=3.5in
\epsffile{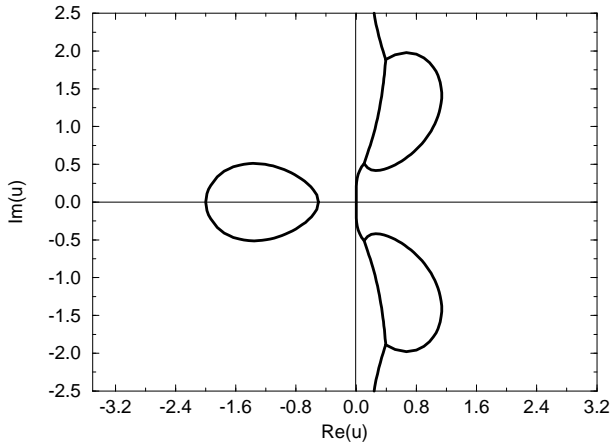}
\caption{Phase diagram of the 1D $Z_6$ clock model in the complex $u$ 
plane. Phases are (a) $R_2$ in the $Re(u) > 0$ half-plane, extending in from 
the far right; (b) $R_{34}$ occupying the truncated lobe-like region in the 
northeast quadrant, and its complex conjugate phase, $R_{34}^*$; 
(c) $R_1$ occupying most of the $Re(u) < 0$ region, and (d) the oval-like
phase $R_{56}$ centered on the negative real $u$ axis.} 
\label{fig1}
\end{figure}

   We find the complex-temperature phase diagram shown in Fig. 1, consisting 
of five different phases.  In the caption, each phase is labelled by the 
eigenvalue (or identically equal eigenvalues) which is (are) dominant within 
it. The first phase is the  
complex-temperature extension (CTE) of the physical $Z_6$-symmetric,
paramagnetic (PM) phase, denoted $R_2$ since $\lambda_2$ is dominant in this
phase.  The other four regions have no overlap with any physical phase and are
thus of O (``other'') type in our previous notation \cite{chisq}.  These 
include two phases $R_{34}$ and $R_{34}^*$ in which $\lambda_{34}$ is dominant;
these are related to each other by complex conjugation and occupy regions 
shaped like truncated lobes in the $Re(u) > 0$ half-plane.  Next, there is the 
phase $R_1$ which occupies most of the left-hand plane $Re(u) < 0$ together 
with a narrow portion to the right of the vertical axis $Re(u) = 0$.  The 
fifth phase, $R_{56}$ comprises an oval-like region centered on
the negative real $u$ axis in which $\lambda_{56}$ is dominant.  A striking
feature of the CT phase diagram is the presence of four intersection points on
the CT boundary ${\cal B}$, at $u \simeq 0.11 + 0.51i$ (given more precisely
below), its inverse, and their complex conjugates.  
This is quite different from the CT phase diagram for the
1D higher-spin Ising model \cite{is1d}, where the curves (and, for 
half-integral spin, the semi-infinite line segment on the negative real $u_s$ 
axis) only meet at the origin of the complex $u_s$ plane, i.e., at $K=\infty$, 
where the model has an FM critical point and a formal essential singularity.
This intersection point has $n_c=4s^2$.  (Given the $u \to 1/u$ 
in the latter model, in the $u_s^{-1}$ plane, the curves also meet at the AFM
critical point at $u_s^{-1}=0$.)  A further difference is the presence in the
$Z_6$ model of CT phases of finite extent in the $u$ plane, viz., $R_{34}$,
$R_{34}^*$, and $R_{56}$.  In the 1D spin $s$ Ising model, all CT phases
extend from the wedge which they occupy in the vicinity of the origin outward
to the circle at infinity in the $u_s$ plane.  A third comparison may be made
with the CT phase diagram for the 1D $q$-state Potts model, which has no 
intersection points on ${\cal B}$ and, for $q \ge 3$, just two phases in the 
$e^{-K}$ plane, the CT extension of the PM phase, of infinite extent, and an 
O phase of finite extent, bounded by the circle $e^{-K}=(-1+e^{i\omega})/(q-2)$
\cite{is1d}.  (The $q=2$ Potts model is equivalent to the spin 1/2 Ising
model.)  

   To discuss the CT phase diagram of the 1D $Z_6$ model, we consider first the
oval boundary of the phase $R_{56}$.  
This is the simplest portion of ${\cal B}$ because it does not contain
any intersection points.  It is the locus of solutions to the degeneracy
condition $|\lambda_1|=|\lambda_{56}|$ where these alternate as the dominant
eigenvalues.  Let $u=re^{i\theta}$. Then
\beq
|\lambda_1|=|\lambda_{56}| \  \Longrightarrow \ 
r\Bigl [ r + 2(1+r^2)\cos\theta + 4r\cos 2\theta \Bigr ] = 0
\label{lam1eqlam5}
\eeq
Note that in obtaining the equation in $r$ and $\theta$ above from the 
condition $|\lambda_1|=|\lambda_{56}|$, we have removed the common factor
$|1-u|^2$, which is irrelevant for ${\cal B}$, since at $u=1$ the condition 
that $\lambda_1$ and $\lambda_{56}$ are dominant is not satisfied (indeed, 
both vanish there).  Further, in eq. (\ref{lam1eqlam5}), the solution $r=0$, 
i.e. $u=0$, is not relevant for ${\cal B}$, since $\lambda_{56}$ is not a
dominant eigenvalue at this point. Equating the remaining factor to zero in 
the region where $\lambda_1$ and $\lambda_{56}$ alternate as dominant 
eigenvalues, one obtains the oval curve shown in Fig. 1. (If one relaxes the 
condition that $\lambda_1$ and $\lambda_{56}$ be dominant eigenvalues,
then one obtains also a curve in the $Re(u) \ge 0$ half-plane passing through
$u=0$, extending in ``northeast'' and ``southeast'' directions; however, this
is not relevant for ${\cal B}$.) Equating the factor in
square brackets in eq. (\ref{lam1eqlam5}) to zero and solving, one gets 
\beq
\cos \theta_{1-56,\pm} = \frac{1}{8}\biggl [ -(r^{-1}+r) \pm 
\sqrt{r^{-2}+26+r^2} \ \ \biggr ]
\label{ctheta15}
\eeq
Observe the $r \to 1/r$ symmetry in (\ref{ctheta15}), in accord with
(\ref{usym}). 
The $-$ sign gives the oval while the $+$ sign gives the above-mentioned curve
in the $Re(u) \ge 0$ region, which does not contribute to ${\cal B}$. Setting 
$\theta=\pi$ in (\ref{lam1eqlam5}) yields the solutions $r=1/2$ and $r=2$, 
i.e. $u=-1/2$ and $u=-2$; these are the 
points where the oval boundary crosses the negative real $u$ axis. 
Setting $r=1$ in the $-$ case of eq. (\ref{ctheta15}) yields the values of 
$\theta$ at which the oval boundary crosses the unit circle $|u|=1$, which 
are given by $\pm \theta_o$, where 
\beq
\theta_{o} = \arccos \biggl ( -\frac{1+\sqrt{7}}{4}\biggr ) = 155.705^\circ
\label{thetacross}
\eeq
The corresponding points are
\beq
\frac{1}{4} \Bigl [ -(1+\sqrt{7}) \pm (8-2\sqrt{7})^{1/2} \Bigr ] \ 
\simeq -0.911 \pm 0.411i 
\label{ovalcross}
\eeq
As one travels along the boundary curve of the oval, $\theta$ starts at the
value $\pi$ at $u=-1/2$, decreases to the value $\theta_o$ as $u$ crosses the
unit circle, and then increases back to $\theta=\pi$ at $u=-2$, and similarly
with the complex conjugate points. 

  The portion of ${\cal B}$ in the $Re(u) \ge 0$ half-plane is more complicated
and consists of several parts, each of which represents a degeneracy condition
of the form $|\lambda_i|=|\lambda_j|$ where $\lambda_i$ and $\lambda_j$
alternate as dominant eigenvalues.  The part of ${\cal B}$ near the origin 
(and hence also, given the symmetry (\ref{usym}), the part farthest out from 
the origin) is a curve consisting of the solution to the condition 
\beq
|\lambda_1|=|\lambda_2| \ \Longrightarrow \ r\Bigl [ 2(1+r^6)\cos \theta 
+ 4r^3\cos 2\theta + r^3\cos 4\theta \Bigr ] = 0
\label{lam1eqlam2}
\eeq
where $\lambda_1$ and $\lambda_2$ are dominant eigenvalues. 
For $r \ne 0$, we analyze the second factor in square brackets.  For small $r$
this reduces to the condition that $\cos \theta=0$, i.e., 
$\theta = \pm \pi/2$, so that this curve passes through the origin in a
vertical direction.  (Equation (\ref{lam1eqlam2}) also has a solution
consisting of a small oval which crosses the negative real $u$ axis at 
$-2^{\pm 1/3}$, but this is not relevant to ${\cal B}$ since $\lambda_1$ and
$\lambda_2$ do not alternate as dominant eigenvalues in this region.) 

The left-hand boundaries of the lobe regions $R_{34}$ and $R_{34}^*$ are 
given by the solution to the condition 
\beq
|\lambda_1|=|\lambda_{34}| \ \Longrightarrow \ 
r\Bigl [ r - 2(1+r^2)\cos \theta \Bigr ] = 0
\label{lam1eqlam3}
\eeq
i.e., $\cos \theta = r/[2(1+r^2)]$, where these eigenvalues are dominant. 
The corresponding curve crosses the unit circle at
$\theta= \arccos(1/4) \simeq \pm 75.52^\circ$.

   The right-hand boundaries of these lobe regions $R_{34}$ and $R_{34}^*$ are 
given by the solution to the condition
\beq
|\lambda_2|=|\lambda_{34}| \ \Longrightarrow \ 
r\Bigl [ 3r(1+r^4) + 2(1+r^6)\cos\theta + 10r^3 \cos 2\theta 
+ 6r^2(1+r^2)\cos 3\theta + 4r^3 \cos 4\theta \Bigr ] = 0
\label{lam2eqlam3}
\eeq
where $\lambda_2$ and $\lambda_{34}$ alternate as dominant eigenvalues.  This
curve crosses the unit circle at $\cos \theta=a^{1/3}+(3/8)a^{-1/3}-1/2$ where
$a=(9+3\sqrt{3})/32$, i.e., $\theta=\pm 41.03^\circ$. 

   As noted above, there are four intersection points, each with $n_c=3$, on 
${\cal B}$.  We denote the one closest to
the origin in the northeast quadrant as $u_t=r_te^{i\theta_t}$; 
the others are then $1/u_t$, and
their complex conjugates, $u_t^*$ and $1/u_t^*$.  The point $u_t$ occurs where
the three eigenvalues $\lambda_1$, $\lambda_2$, and $\lambda_{34}$ are all
degenerate and dominant, i.e., where the phase boundaries defined by the 
conditions $|\lambda_1|=|\lambda_2|$, $|\lambda_1|=|\lambda_{34}|$, and 
$|\lambda_2|=|\lambda_{34}|$ all meet.  This point is given by the solution 
$r_t=0.518684...$ to the equation
\beq
2(r^{-6}+r^6)-18(r^{-2}+r^2)-31=0
\label{tripr}
\eeq
This equation is manifestly symmetric under the symmetry (\ref{usym}), and has
two positive real reciprocal roots, of which the smaller is the one which we
have listed above.  The corresponding angle is $\theta_t=78.207851^\circ$, so
that 
\beq
u_t \simeq 0.1059993 + 0.5077375i
\label{utvalue}
\eeq

   A general characteristic of 1D spin models with short-range forces is that
the points $K = \pm \infty$ are critical points, so that in the present case, 
the phase boundary ${\cal B}$ passes through the respective origins in the $u$
and $u^{-1}$ planes.  This is reminiscent of how ${\cal B}$ passed through the
origin of the $u_s$ plane for the 1D spin $s$ Ising model and the $e^{-K}$
plane for the 1D $q$-state Potts model \cite{is1d}.

\begin{figure}
\epsfxsize=3.5in
\epsffile{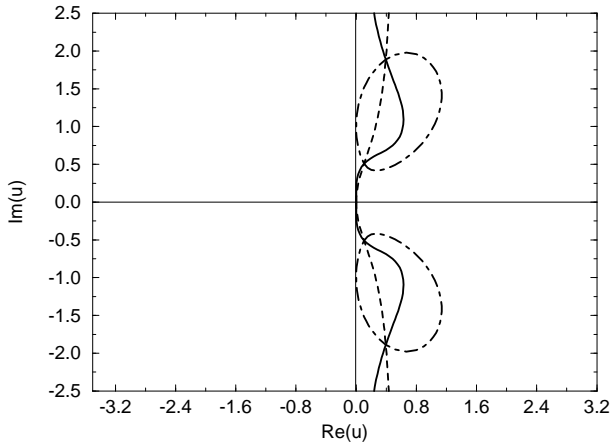}
\caption{Solutions with $Re(u) \ge 0$ of the conditions 
(a) $|\lambda_1|=|\lambda_2|$ (solid curve); (b) $|\lambda_1|=|\lambda_{34}|$
(dashed curve); and (c) $|\lambda_2|=|\lambda_{34}|$ (dot-dashed curve).}
\label{fig2}
\end{figure}

   In order to gain further insight into the components of ${\cal B}$, 
one may determine the full locus of points comprising the solution of the 
degeneracy condition $|\lambda_i|=|\lambda_j|$ without the constraint that
$\lambda_i$ and $\lambda_j$ be dominant eigenvalues.  One can then see how
subsets of this locus form the various components of ${\cal B}$.  In Fig. 2 we
show the portions of the solutions to the $|\lambda_i|=|\lambda_j|$ equations 
for $(i,j)=(1,2)$, (1,34), and (2,34) which are involved with the finite-$K$
intersection points.  The loci of solutions to $|\lambda_2|=|\lambda_{34}|$
and $|\lambda_1|=|\lambda_{56}|$ also contain curves which pass through the
origins of the $u$ and $u^{-1}$ planes; we omit these from Fig. 2 since they do
not contribute to ${\cal B}$ except at the respective origins of the $u$ and 
$u^{-1}$ planes, which are the zero-temperature FM and AFM critical points of 
the model. The intersection point $u_t$, together with its inverse and their 
complex conjugates, are seen to be the (finite-$K$) points at which all of 
the curves $|\lambda_i|=|\lambda_j|$ for $(i,j)=(1,2)$, (1,34), and (2,34) 
cross.  While a solution curve of $|\lambda_i|=|\lambda_j|$ for a
given $(i,j)$ continues smoothly through $u_t$, the actual CT phase boundary 
${\cal B}$ has a discontinuous tangent there and heads off in a different
direction as one moves away from $u_t$, since it is then determined by a
different degeneracy condition $|\lambda_j|=|\lambda_k|$. 

   We generalize our results as follows: for a 1D discrete spin model with 
short-ranged interactions, so that the transfer matrix has a finite number of
eigenvalues, the CT phase boundary ${\cal B}$ has an intersection point with
$n_c$ curves meeting if and only if $n_c$ (in general distinct) dominant 
eigenvalues become degenerate in magnitude at this point.  

   Of course, there are differences between $d=1$ and $d \ge 2$ models. 
For $d \ge 2$, ${\cal T}$ has an infinite number of eigenvalues, rather than 
the finite number which it has for $d=1$.  
Further, since $d > 1$ is above the lower critical
dimensionality $d_{\ell. c. d.}=1$ of a usual discrete spin model, two
eigenvalues which are distinct in the symmetric phase may be degenerate in the
phase with spontaneously broken symmetry and ferromagnetic (FM) or
antiferromagnetic (AFM) long-range order, again in contrast to the situation 
in $d=1$, where there is no symmetry-breaking phase for finite $K$. 
However, these differences do not prevent the 1D $Z_6$ model from having the
same feature, viz., an intersection point on ${\cal B}$ with $n_c=3$, as 
higher-dimensional models such as the spin 1/2 3D Ising model. 

   Indeed, we have previouly pointed out some intriguing connections between 
certain features of the respective CT phase boundaries of the higher-spin 
Ising model in 1D and 2D \cite{hs,is1d}.  These included the observation that 
(exactly known) angles at which phase boundaries crossed the 
unit circle in the 1D case were consistent with being equal to values of
analogous angles in the 2D model. Remarkably, we find the same kind of 
connection here.  The physical phase structure of the 2D $Z_6$ model consists
of (i) a disordered, $Z_6$-symmetric high-temperature phase, (ii) a 
$Z_6$-symmetric intermediate-temperature phase with algebraic decay of 
correlations, and (iii) a broken-symmetry low-temperature phase with FM (AFM)
long-range order for $J > 0$ ($J < 0$) \cite{znpd}.  From the CT zeros of the 
partition function for the $Z_6$ model calculated on a $6 \times 7$ 
lattice \cite{martin}, which we have confirmed in an independent calculation, 
one infers that in the ``northwest'' and ``southwest'' regions of the $u$ 
plane, in the thermodynamic limit, ${\cal B}$ has intersection points with 
$n_c=4$, $n_t=2$, and $m=2$ at $e^{\pm i\theta_o}$, where $\theta_o$ is
the same angle as we have calculated in eq. (\ref{thetacross}).  This 
suggests that one can gain valuable information about certain features of the 
complex-temperature properties of a higher-dimensional spin model from the CT 
properties of the corresponding 1D model, reminiscent of the fact that 
$\epsilon$ expansions starting from $d=d_{\ell. c. d.}$ gave similar insight 
into the physical phase transition in models for 
$d > d_{\ell. c. d.}$ \cite{eps}.  

This research was supported in part by the NSF grant PHY-93-09888.

\vfill
\eject
\end{document}